\documentclass[aps,prl,twocolumn,superscriptaddress,10pt]{revtex4-2}

\usepackage{graphicx}
\usepackage{amsmath,amssymb,bm}
\usepackage{siunitx}
\usepackage{hyperref}

\begin{document}

\title{High-throughput, high-brightness, ultrashort 90~keV electrons at 40 kHz}

\author{K.~Amini}
\email{Corresponding author: kasra.amini@mbi-berlin.de} 
\affiliation{Max-Born-Institut, Max-Born-Str.~2A, 12489 Berlin, Germany}

\author{T.C.H.~de~Raadt}
\affiliation{Doctor X Works, Eindhoven, The Netherlands}

\author{J.G.H.~Franssen}
\affiliation{Doctor X Works, Eindhoven, The Netherlands}

\author{B.~Siwick}
\affiliation{Department of Physics, McGill University, Montreal, Canada}

\author{O.J.~Luiten}
\affiliation{Department of Applied Physics, Eindhoven University of Technology, Eindhoven, The Netherlands}

\author{A.~Ryabov}
\affiliation{Max-Born-Institut, Max-Born-Str.~2A, 12489 Berlin, Germany}
\date{\today}

\begin{abstract}
Radiofrequency-compressed keV electron sources for ultrafast electron diffraction (UED) face competing demands: short pulses require low charge, yet weak scatterers demand high flux; high repetition rates enable signal averaging, yet most systems operate $\lesssim$1 kHz with low detection efficiency. Here, we demonstrate a 90 keV DC-RF source operating at 40 kHz with direct electron detection that address these challenges simultaneously. THz streaking retrieves compressed pulse durations of 97 ± 3 fs (FWHM) at 370 aC and 114 ± 47 fs (FWHM) at 2.8 fC. Long-term $t_0$ timing drifts, characterized independently both by convolution analysis of compression data and direct THz streaking measurements, lie between 65 -- 95 fs (FWHM), among the lowest reported for RF-based systems. At low charge (17 aC), we report an intrinsic pulse duration of 56 fs (FWHM) from comparison of simulations to measured compression data, among the shortest for keV UED at $>$16 aC. Moreover, 2.8 fC bunches, combined with 40 kHz repetition rate and direct detection, produce a detectable normalized throughput that is one (three-to-four) orders of magnitude higher than existing keV (MeV) sources. This enables practical UED studies of weakly scattering samples and processes previously impractical due to low cross-sections and long acquisition times.
\end{abstract}

\maketitle
\begin{figure*}[t!]
  \centering
  \includegraphics[width=0.80\linewidth]{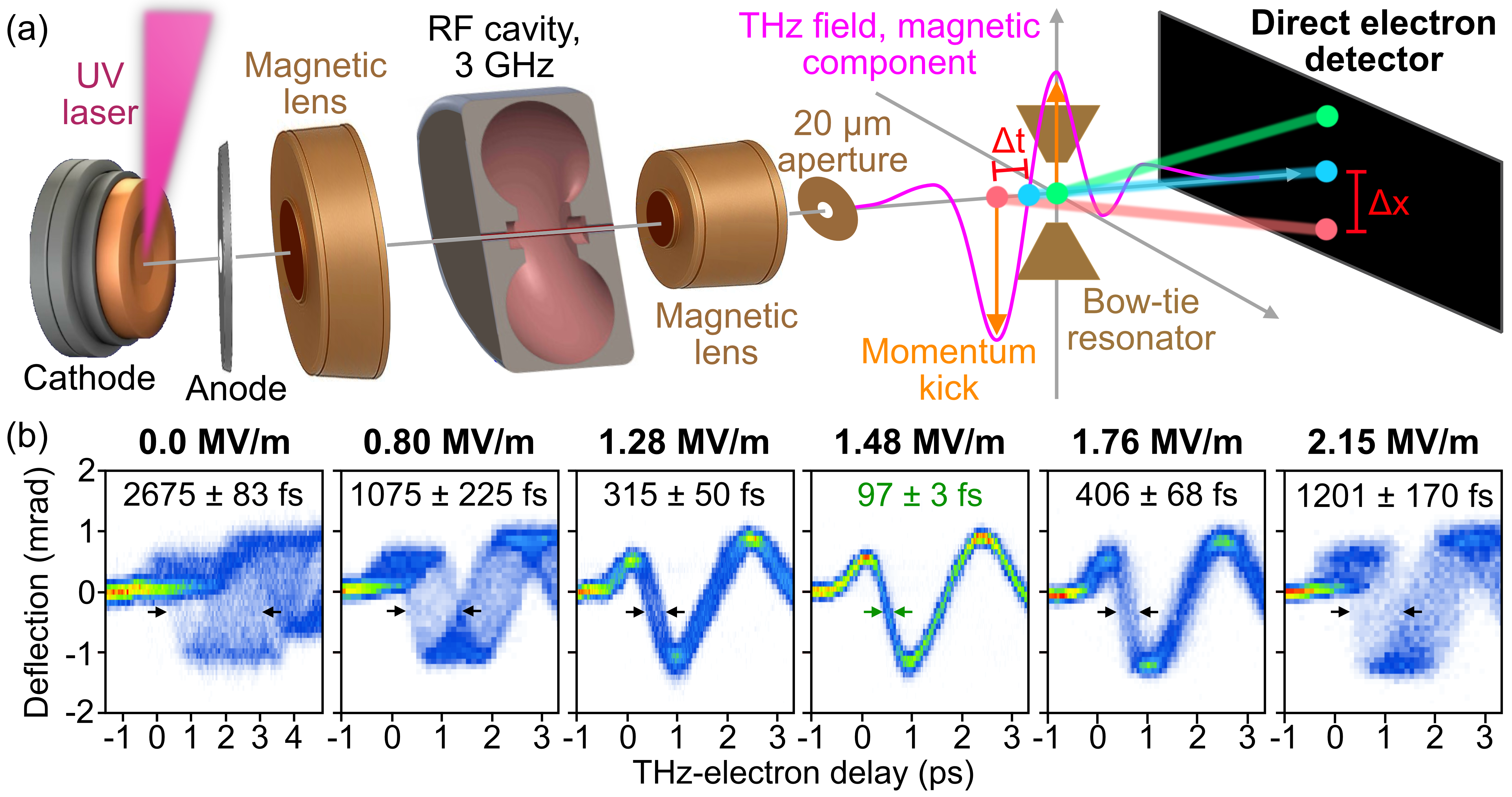}
  \caption{(a) Schematic of the compact 90~keV UED apparatus operating at 40~kHz. A femtosecond UV pulse photoemits electrons from a direct current (DC) gun, which are transversely collimated, longitudinally compressed in a 3 GHz RF cavity, and focussed through a 20\,\textmu m aperture and onto a THz bowtie resonator for temporal characterization via THz electron streaking with direct electron detection. (b) THz streaking measurement of an electron bunch containing $\sim$2,300 electrons (370~aC), demonstrating compression from 2.7~ps to 97~±~3~fs (FWHM) at an RF field of 1.48~MV/m. Over-compression at higher fields shifts the temporal focus upstream of the resonator, increasing the measured duration. Uncertainties represent the FWHM deviation in the retrieved pulse duration from 3-5 independent measurements.}
  \label{fig:setup}
\end{figure*}
Ultrafast electron diffraction (UED)\cite{Filippetto2022,Centurion2022,Amini2023} directly visualizes atomic motion on femtosecond and Ångström scales, enabling studies to track photochemical bond breaking and rearrangement through conical intersections \cite{Yang2018,Wolf2019,Yang2020,Wang2025,Wang2025a, Jiang2025} to mapping lattice vibrations\cite{Qi2020,Chen2025}, charge-density waves\cite{Kogar2020, Duan2021} and two-dimensional engineering of  quantum materials \cite{Duan2021,Duncan2025}. Yet many systems of chemical and biological interest remain beyond reach: dilute molecular gases with low absorption or scatttering cross-sections, fragile samples that cannot withstand high pump fluences, weakly scattering species that produce signals below current detection thresholds, or dynamics that evolve faster than available typical electron pulse durations. Two of the underlying limitations are the detected electron flux (i.e., throughput), which has received less attention than temporal resolution and beam brightness as a UED performance metric, and achieving the shortest electron pulse possible to capture the fastest dynamics.

These limitations stem from competing constraints in current UED instrumentation. Compact~keV sources suffer from strong space-charge broadening and typically operate at low repetition rates ($\leq$ 1 kHz). Even with radiofrequency (RF) \cite{vanOudheusden2010,Otto2017} and terahertz (THz) \cite{Kealhofer2016} electron compression,~keV beams have so far been constrained to durations of 100-150~fs (FWHM) at $\sim$fC or higher bunch charges, while single-electron operation at 25--50 kHz achieves durations of tens of femtoseconds but sacrifices throughput by orders of magnitude \cite{Volkov2021,Tauchert2022}. MeV facilities\cite{Shen2019, Hennicke2025}, although capable of generating 8-29~fs (FWHM) electron pulses\cite{Qi2020,Yang2025}, operate at $<$1 kHz repetition rates and rely on indirect detection schemes. As a result, no existing UED source simultaneously delivers short pulse duration, high brightness, and high detected flux for capturing ultrafast dynamics in weakly scattering, photoexcited samples.

To quantify this trade-off, we define the detectable, normalized five-dimensional (5D) throughput,
\begin{equation}
T_{\mathrm{np,det}} = f_{\mathrm{rep}}\, \cdot \mathrm{DQE}\, \cdot B_{\mathrm{np}},
\end{equation}\label{Eq1}
\noindent where $f_{\mathrm{rep}}$ is the repetition rate (electrons per second), DQE is the detective quantum efficiency (fraction detected), and $B_{\mathrm{np}}$ is the 5D normalized brightness (electrons per phase-space volume per pulse) given by,
\begin{equation}
B_{np}
= \frac{Q}{\sqrt{2}\,\pi^{5/2}\,\epsilon_n^2\,\sigma_t}.
\label{Eq:5D_brightness}
\end{equation}
\noindent where $Q$ is the bunch charge, $\epsilon_{n}$ is the normalized emittance, and $\sigma_{\mathrm{t}}$ the root mean square (rms) bunch duration. $T_{\mathrm{np,det}}$ therefore quantifies the number of detected electrons per second within a defined phase-space volume, which is the relevant metric for signal accumulation in weak-scattering experiments. State-of-the-art UED sources achieve \(T_{\mathrm{np,det}} \lesssim 10^{16}\) A/m\(^2\)/rad\(^2\)/s, sufficient for the study of systems with large absorption and scattering cross-sections but inadequate for dilute or weakly scattering targets, which require a throughput improvement of at least one-to-two orders of magnitude.

Here, we demonstrate a compact 90~keV RF-compressed source operating at 40 kHz that achieves $T_{\mathrm{np,det}} = 3.7\times10^{18}\ \mathrm{A/m^{2}/rad^{2}/s}$ which is one (three-to-four) orders of magnitude higher than existing~keV (MeV) sources. At low bunch charge (17~aC), we achieve intrinsic pulse durations of 56~fs (FWHM), among one of the shortest reported for~keV UED; at moderate (370~aC) and high (2.8~fC) charge, pulse durations of 97~fs and 138~fs (FWHM), respectively, maintain the highest throughput. Combined with direct electron detection (DQE = 0.85), 6~fs short-term jitter, and $\sim$100~fs long-term timing drifts, this establishes a new operating regime for future studies of dilute, fragile, and weakly scattering systems previously beyond reach.

\begin{figure*}[t!]
  \centering
  \includegraphics[width=0.90\linewidth]{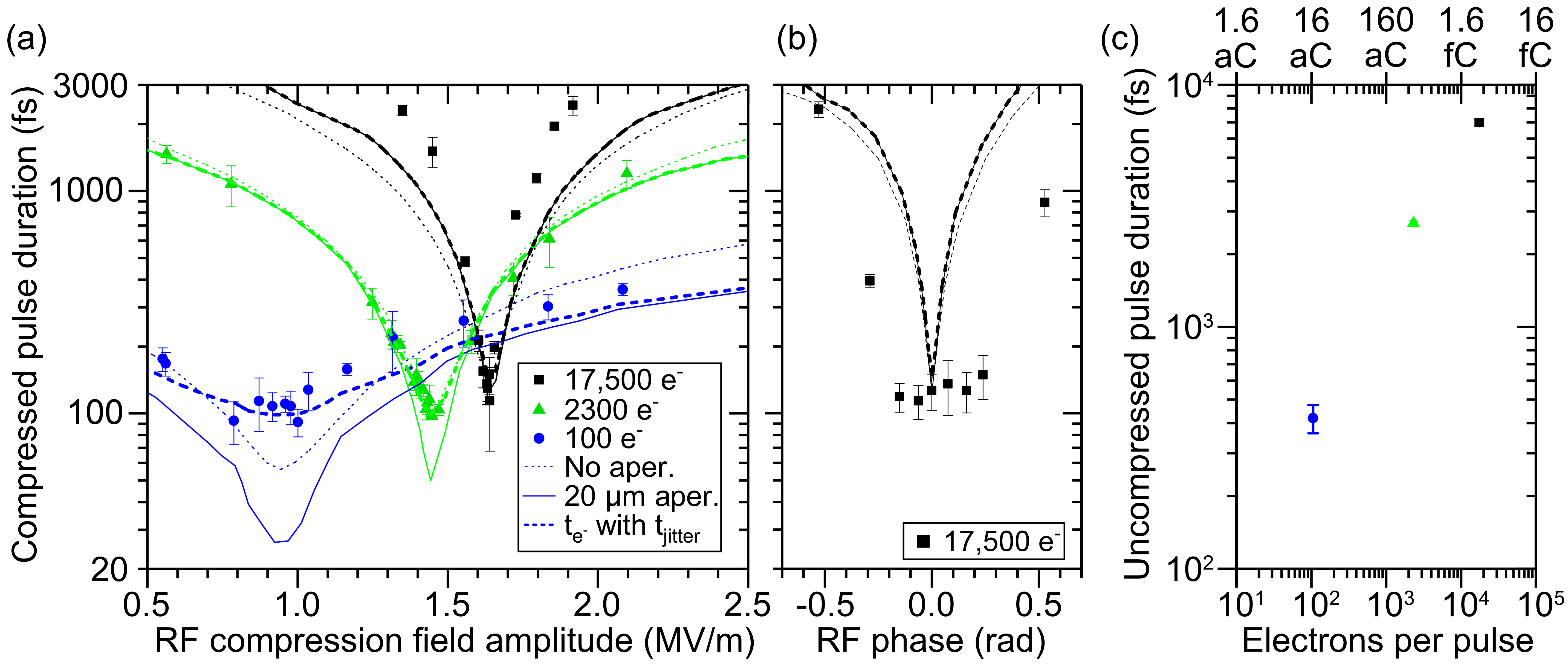}
  \caption{(a) Measured (symbols) and GPT simulated (dashed lines) electron pulse duration as a function of RF field amplitude. Uncertainties represent the FWHM deviation in the retrieved pulse duration from 3-5 independent measurements. (b) Dependence of compressed electron pulse duration on RF phase for a 2.8~fC bunch. (c) Measured uncompressed electron pulse duration as a function of electrons per pulse. All values are given in FWHM. Horizontal error bars are smaller than the symbol size.}
  \label{fig:RF_scans}
\end{figure*}

Figure \ref{fig:setup}a shows our experimental setup. A 90~fs ultraviolet (UV) pulse\cite{Ryabov2025} photoemits electrons from a copper cathode biased at 90~keV in a direct current (DC) electron accelerator with an accelerating field strength of 10~MV/m. The electron beam is transversely collimated by a magnetic lens, longitudinally compressed in a 3 GHz RF cavity, and transversely focused by a second magnetic lens through a 20\,\textmu m aperture onto a THz bowtie resonator. The aperture selects the highest-brightness core of the beam while allowing for a sufficiently reduced beam diameter to pass through the clear aperture of the resonator. A 0.5 THz pulse\cite{Ryabov2025}, resonantly enhanced at the resonator, deflects electrons according to their arrival time, mapping the bunch longitudinal (i.e., temporal) distribution onto the transverse plane of a direct electron detector. Figure \ref{fig:setup}b shows THz electron streaking measurements for $\sim$2,300 electrons per pulse (i.e., 370~aC). The uncompressed beam has a duration of 2.7 ± 0.1~ps (FWHM) and is compressed to 97~fs (FWHM) at an RF field of 1.48~MV/m, with~±~3~fs reproducibility across three independent scans. At higher fields ($>$2~MV/m), over-compression of the electron bunch shifts the temporal focus upstream before the bowtie resonator, increasing the measured pulse duration at the resonator. The streaking calibration from Fig. \ref{fig:setup}b (23~fs/pixel, 3.0\,\textmu mrad/fs deflection) gives an estimated THz field strength of $\sim$4.7 kV/cm\cite{Kealhofer2016}. We first characterize compression at moderate charge (370~aC) before exploring the full dynamic range in Fig.~\ref{fig:RF_scans}.

Figure \ref{fig:RF_scans}a shows the retrieved FWHM pulse duration as a function of RF field amplitude. For a 370~aC bunch (green triangles), we observe characteristic under- and over-compression behavior. General Particle Tracer (GPT\cite{GPT}) simulations with the 20\,\textmu m aperture (green solid) show good agreement with measurements down to $\sim$200~fs, below which a systematic deviation emerges: at optimal compression, GPT predicts 68~fs (FWHM) duration with the aperture, yet we retrieve a measured total duration of 97~±~3~fs (FWHM). This discrepancy is due to a timing jitter of $\sim$85~fs. Convolving the GPT-simulated, aperture-cut durations with 85~fs jitter (green dashed) reproduces the measured compression curve across the full range of RF field amplitudes. GPT simulations of the full beam without aperture (green dotted) predict a 97~fs (FWHM) duration at optimal compression.

We then study RF compression at higher and lower bunch charges. At high charge (2.8~fC, $\sim$17,500~electrons/pulse), severe space-charge broadening produces an uncompressed pulse of 7.0~±~0.2~ps (see Fig. \ref{fig:RF_scans}c), compressed by a factor of 60 to 114~±~47~fs (FWHM; see black squares). GPT simulations with the 20\,\textmu m aperture (black solid) predict 124.6 fs (FWHM) at optimal compression, while GPT simulations without aperture (black dotted) predicts a duration of 138 fs (FWHM). Near the compression point, a timing jitter convolution of $\sim$65 fs (black dashed) is sufficient to fit the upper range of the measured values, indicating that space-charge effects dominate over timing jitter at high charge. At low charge (17~aC, $\sim$100~electrons/pulse, blue circles), weak space-charge forces lead to an uncompressed duration of 420~±~56~fs (FWHM), compressible to 91.5~±~13~fs (FWHM) at $\sim$1.0~MV/m. Here, GPT predicts significantly shorter durations: 26 fs (FWHM) with aperture (blue solid) and 56 fs (FWHM) for the full beam (blue dotted). Convolving the aperture-simulated pulse duration of 26 fs with a 95~fs timing jitter (blue dashed) reproduces the measured compression data (blue circles) across nearly all RF field amplitudes. This yields a total temporal resolution of 98.5~fs which is in good agreement with the retrieved value (91.5~±~13~fs), supporting an intrinsic full-beam pulse duration of 56 fs (FWHM) at 17~aC, which is among the shortest reported for keV UED, with the measured duration limited by timing jitter rather than electron pulse length.

We next investigate the sensitivity of temporal compression to RF phase for a 2.8~fC bunch (Fig. \ref{fig:RF_scans}b). The compression behavior shows similar under- and over-compression trends at negative and positive phase delays, respectively. At the optimal compression point, the electron bunch samples the zero-crossing of the RF field where the slope is maximum; since the second derivative has a value of zero here, compression is inherently insensitive to small phase deviations (±0.16~rad). Away from optimal phase, the bunch experiences net acceleration or deceleration, shifting the electron arrival time at the THz resonator. The measured duration (black squares) agrees with GPT simulations using an aperture (black solid) near the optimal compression. Convolving the GPT-simulated duration with an 83 fs timing jitter (black dashed) remains withing the upper range of the measured error bars. Overall, we compress electron bunches from 0.42 -- 7.0~ps down to 91.5 -- 114~fs (FWHM), corresponding to intrinsic full-beam durations of 56 -- 138 fs (FWHM), across bunch charges from 17~aC to 2.8~fC. We next benchmark the our results against state-of-the-art UED sources.

\begin{figure}[b!]
  \centering
  \includegraphics[width=\linewidth]{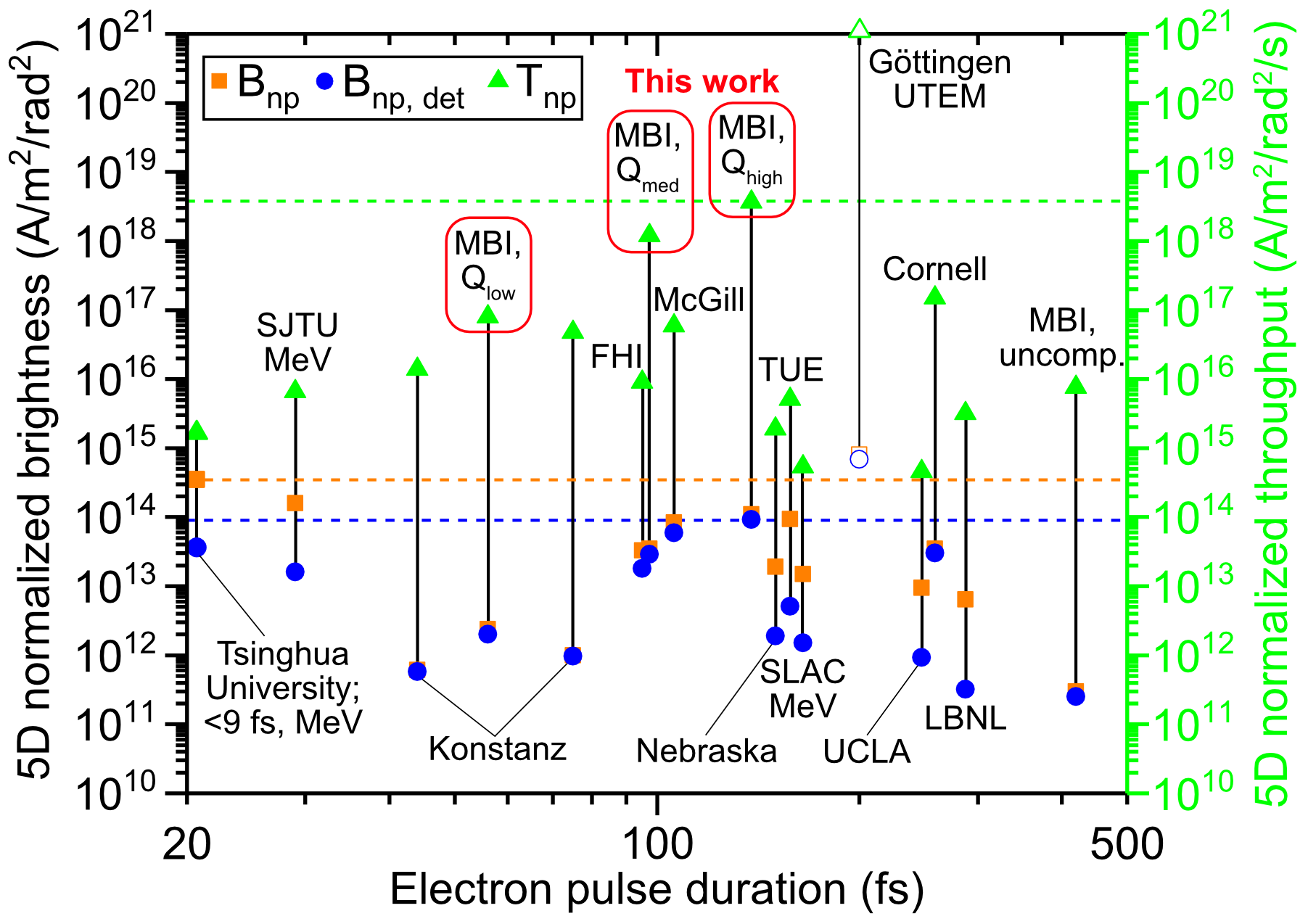}
  \caption{5D normalized brightness $B_{\mathrm{np}}$ (orange squares) as a function of electron pulse duration (FWHM), together with detectable 5D normalized brightness $B_{\mathrm{np,det}}$ (blue circles) accounting for detector efficiency, and detectable 5D normalized throughput $T_{\mathrm{np,det}}$ (green triangles) additionally accounting for the source repetition rate. Open symbols denote UTEM nanotip sources optimized for nanoscale imaging (see main text). The horizontal dashed lines indicate the setup with the highest parameter.}
  \label{fig:Brightness}
\end{figure}

To quantitatively benchmark our source, we consider the 5D normalized brightness given by Eq.~\ref{Eq:5D_brightness} (see orange squares in Fig.~\ref{fig:Brightness}). For brightness estimates, we use the pulse durations from GPT full-beam simulations (without aperture): 138.1~fs (FWHM) at 2.8~fC, 97.4~fs (FWHM) at 370~aC, and 56~fs (FWHM) at 17~aC. Our high-charge (2.8~fC) and moderate-charge (370~aC) modes compare favorably to state-of-the-art instruments, particularly keV systems and the highest-brightness MeV source (dashed orange line). Accounting for detector efficiency, we define a detectable 5D brightness $B_{\mathrm{np,det}} = B_{\mathrm{np}} \times \mathrm{DQE}$ (blue circles), where DQE is the detective quantum efficiency relating the signal-to-noise ratio (SNR) before and after detection.

For many MeV facilities, $B_{\mathrm{np,det}}$ is significantly lower than $B_{\mathrm{np}}$ due to the low DQE of lens-coupled scintillators with electron-multiplying charge-coupled devices (EMCCDs; DQE~$\sim$0.1 at 0.1 -- 3.7 MeV). Fiber-coupled scintillators with complementary metal-oxide–semiconductor (CMOS) sensors improve this to $\sim$0.55 in integrating mode at 100~keV beams, and single-electron counting can approach unity but only at $\ll 1$~electron/pixel/frame \cite{Kealhofer2015}. By contrast, our direct electron detection (DQE = 0.85 at 100~keV\cite{FernandezPerez2021}) preserves 85\% of the beam brightness, yielding $B_{\mathrm{np,det}} \approx B_{\mathrm{np}}$ and placing our source among the highest $B_{\mathrm{np,det}}$ reported (blue dashed line). Combined with our 40~kHz repetition rate, this produces a detectable 5D normalized throughput $T_{\mathrm{np,det}} = 3.7\times10^{18}\ \mathrm{A/m^{2}/rad^{2}/s}$ (green triangles; see dashed green line), which is over one order of magnitude higher than existing~keV sources and three-to-four orders of magnitude higher than MeV facilities. We note that ultrafast transmission electron microscope (UTEM) nanotip sources\cite{Feist2017,Domroese2025} achieve exceptional normalized brightness ($B_{\mathrm{np}} \sim 10^{14}$--$10^{15}$~$\mathrm{A/m^{2}/rad^{2}}$) and throughput ($T_{\mathrm{np,det}} \sim 10^{21}$~$\mathrm{A/m^{2}/rad^{2}/s}$) through pm$\cdot$rad emittances and 2~MHz repetition rates, respectively. However, these sources operate in a fundamentally different regime ($\sim$0.5 electron/pulse, 200~fs) optimized for nanoscale imaging and diffraction rather than diffraction from extended, weakly scattering samples. For example, gas-phase UED requires a minimum electron flux of $\sim10^{7}$ electrons/s across 100~\textmu m -- mm lengths for sufficient gas-phase scattering signals.

\begin{figure}[t!]
  \centering
  \includegraphics[width=0.80\linewidth]{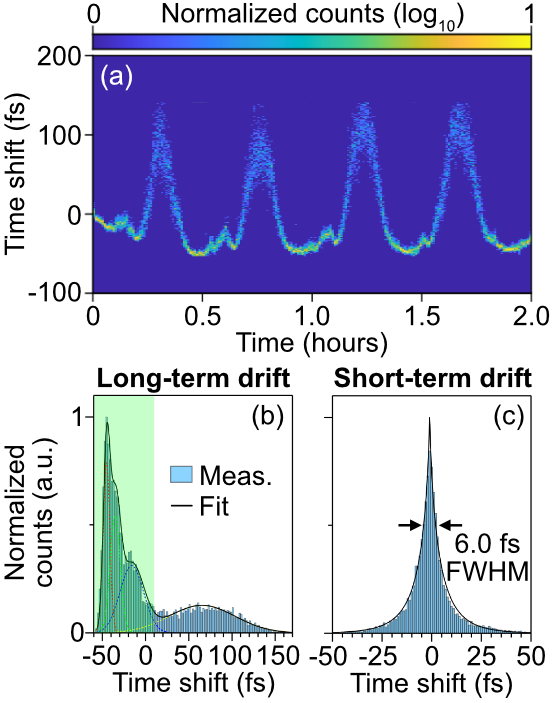}
  \caption{Timing stability characterization via THz streaking. (a) Measured RF-laser timing shift as a function of acquisition time. (b-c) Corresponding histogram analysis of long-term and short-term timing shift data from panel (a). Green shaded shows that 70\% of data is confined to a 60~fs window by the active RF-laser synchronization system.}
  \label{fig:TimingStability}
\end{figure}

While high repetition rate enables throughput, timing stability determines whether this flux is usable for pump-probe experiments. We directly measure the timing stability of the electron source using THz streaking (see Fig.~\ref{fig:TimingStability}). By positioning at the steepest slope of the THz deflectogram, midway between the first and second extrema (see arrows in Fig. \ref{fig:setup}b), we maximize sensitivity to arrival time variations. While maintaining this fixed THz-electron delay, we monitor electron arrival time fluctuations and compare with the 65 -- 95~fs (FWHM) jitter inferred from the convolution of the aperture-cut GPT data in Fig. \ref{fig:RF_scans}a-b discussed earlier. Over two hours, sufficient for a complete gas-phase UED scan at 40~kHz, we observe a long-term $t_{\mathrm{0}}$ drift of 60–120~fs (FWHM), with 70\% of data confined within a 60~fs window (green shaded area in Fig.~\ref{fig:TimingStability}b) by our active RF–laser synchronization system based on Ref.\,\cite{Otto2017}. The short-term timing jitter of 6~fs FWHM (Figs.~\ref{fig:TimingStability}c) is limited by the RF-laser phase lock, while the longer-term $t_{\mathrm{0}}$ drift reflects environmental variations affecting the RF cavity.

In summary, we have demonstrated a compact DC–RF~keV UED source operating at 40~kHz employing an active RF–laser synchronization system with two key results. First, we achieve intrinsic pulse durations of 56~fs (FWHM) at 17~aC, among the one of the shortest reported for keV UED, validated by the agreement between GPT simulations and measured RF compression data convolved with a timing jitter of 95~fs. THz streaking measurements of the long-term $t_0$ drift confirm values between 60 -- 100~fs (FWHM), with short-term timing jitter measured as 6 fs (FWHM). Nonetheless, the timing jitter currently limits the measurable retrieved pulse durations from THz electron streaking measurements to approximately 100 fs, preventing the full utilization of the 56 fs intrinsic pulse duration. Future work will aim to reduce this jitter below 50 fs through improved RF-laser synchronization. At the 20~fs timing jitter level of our optical timing jitter, THz streaking would retrieve a total value of 59~fs, closely approaching the 56~fs intrinsic pulse duration. Second, at high bunch charge (2.8~fC, 138~fs), we obtain a detectable 5D normalized throughput of $T_{\mathrm{np,det}} = 3.7\times10^{18}\ \mathrm{A/m^{2}/rad^{2}/s}$, which is an order of magnitude higher than existing~keV UED sources and three-to-four orders higher than MeV sources. Even at moderate bunch charge (370~aC, 97~fs), our source maintains the highest throughput among traditional UED systems.

These capabilities enable a new operating regime for UED studies of gas-phase and other weakly scattering samples and processes (e.g., inelastic scattering). At comparable bunch charge ($\sim$3~fC) to MeV sources at the sample plane, the combination of a higher repetition rate ($100\times$) and larger scattering cross-section ($2\times$) at 90 keV can yield acquisition times down to 6 minutes, which is significantly faster than the typical 20-hour acquisition time of gas-phase scans at MeV facilities. The projected acquisition time is also faster than the 30-minute period of long-term $t_0$ drift observed in our timing stability measurements, effectively outrunning environmental timing instabilities. This will enable systematic studies of molecules with small photoabsorption or scattering cross-sections, inelastic scattering processes, and wavelength-dependent photochemistry, which have previously been impractical due to the acquisition time constraints. Future work will demonstrate these capabilities on benchmark gas-phase photochemical systems.

\begin{acknowledgments}
We acknowledge financial support from the European Research Council for ERC Starting Grant ``TERES'' (Grant No. 101165245) and Lasers4EU (Grant No. 101131771, Project ID 37011). We are grateful to Peter Baum, Fernando Antonio Rodr\'{\i}guez D\'{\i}az, Arnaud Rouz\`ee, Ingo Will, Martin Otto, Eamon Egan, Roman Peslin, Wolfgang Krueger, Thomas Mueller, Johannes Tuemller and Marc J. J. Vrakking for support. 
\end{acknowledgments}

\bibliographystyle{apsrev4-2}
\bibliography{refs}

@article{Domroese2025,
  title={Megahertz cycling of ultrafast structural dynamics enabled by nanosecond thermal dissipation},
  author={Domr{\"o}se, Till and da Camara Silva, Leonardo and Ropers, Claus},
  journal={Applied Physics Letters},
  volume={126},
  number={12},
  year={2025},
  publisher={AIP Publishing}
}

@article{Feist2017,
  title={Ultrafast transmission electron microscopy using a laser-driven field emitter: Femtosecond resolution with a high coherence electron beam},
  author={Feist, Armin and Bach, Nora and da Silva, Nara Rubiano and Danz, Thomas and M{\"o}ller, Marcel and Priebe, Katharina E and Domr{\"o}se, Till and Gatzmann, J Gregor and Rost, Stefan and Schauss, Jakob and others},
  journal={Ultramicroscopy},
  volume={176},
  pages={63--73},
  year={2017},
  publisher={Elsevier}
}

@article{Tauchert2022,
  title={Polarized phonons carry angular momentum in ultrafast demagnetization},
  author={Tauchert, Sonja R and Volkov, Mikhail and Ehberger, Dominik and Kazenwadel, D and Evers, Martin and Lange, Hannah and Donges, Andreas and Book, Alexander and Kreuzpaintner, W and Nowak, U and others},
  journal={Nature},
  volume={602},
  number={7895},
  pages={73--77},
  year={2022},
  publisher={Nature Publishing Group UK London}
}

@article{Volkov2021,
  title={Photo-Switchable Nanoripples in Ti3C2 T x MXene},
  author={Volkov, Mikhail and Willinger, Elena and Kuznetsov, Denis A and Müller, Christoph R and Fedorov, Alexey and Baum, Peter},
  journal={ACS Nano},
  volume={15},
  number={9},
  pages={14071--14079},
  year={2021},
  publisher={ACS Publications}
}

@misc{GPT,
  author = {{Pulsar Physics}},
  title = {{General Particle Tracer (GPT)}},
  howpublished = {\url{https://www.pulsar.nl/gpt/}},
  year = {2025}
}

@article{Filippetto2022,
  title={Ultrafast electron diffraction: Visualizing dynamic states of matter},
  author={Filippetto, Daniele and Musumeci, Pietro and Li, RK and Siwick, Bradley John and Otto, MR and Centurion, Martin and Nunes, JPF},
  journal={Review Modern Physics},
  volume={94},
  number={4},
  pages={045004},
  year={2022},
  publisher={APS}
}

@article{Kogar2020,
  title={Light-induced charge density wave in LaTe3},
  author={Kogar, Anshul and Zong, Alfred and Dolgirev, Pavel E and Shen, Xiaozhe and Straquadine, Joshua and Bie, Ya-Qing and Wang, Xirui and Rohwer, Timm and Tung, I-Cheng and Yang, Yafang and others},
  journal={Nature Physics},
  volume={16},
  number={2},
  pages={159--163},
  year={2020},
  publisher={Nature Publishing Group UK London}
}

@article{Chen2025,
  title={Structural Contribution to Light-Induced Gap Suppression in Ta 2 NiSe 5},
  author={Chen, Zijing and Xu, Chenhang and Xie, Chendi and Tang, Weichen and Liu, Qiaomei and Wu, Dong and Xu, Qing and Jiang, Tao and Zhu, Pengfei and Zou, Xiao and others},
  journal={Physical Review Letters},
  volume={135},
  number={9},
  pages={096901},
  year={2025},
  publisher={APS}
}

@article{Duan2021,
  title={Optical manipulation of electronic dimensionality in a quantum material},
  author={Duan, Shaofeng and Cheng, Yun and Xia, Wei and Yang, Yuanyuan and Xu, Chengyang and Qi, Fengfeng and Huang, Chaozhi and Tang, Tianwei and Guo, Yanfeng and Luo, Weidong and others},
  journal={Nature},
  volume={595},
  number={7866},
  pages={239--244},
  year={2021},
  publisher={Nature Publishing Group UK London}
}

@article{Hennicke2025,
  title={3D atomic structure determination with ultrashort-pulse MeV electron diffraction},
  author={Hennicke, Vincent and Hachmann, Max and Klar, Paul Benjamin and Reinke, Patrick YA and Pakendorf, Tim and Meyer, Jan and Delsim-Hashemi, Hossein and Barthelmess, Miriam and Veedu, Sreevidya Thekku and Fischer, Pontus and others},
  journal={arXiv preprint arXiv:2507.06936},
  year={2025}
}

@article{Shen2019,
  title={Femtosecond gas-phase mega-electron-volt ultrafast electron diffraction},
  author={Shen, Xiaozhe and Nunes, JPF and Yang, J and Jobe, RK and Li, RK and Lin, Ming-Fu and Moore, B and Niebuhr, M and Weathersby, SP and Wolf, TJA and others},
  journal={Structural Dynamics},
  volume={6},
  number={5},
  year={2019},
  pages={054305},
  publisher={AIP Publishing},
  doi={10.1063/1.5120864},
  url={https://doi.org/10.1063/1.5120864}
}

@article{Duncan2025,
  title   = {Photoinduced twist and untwist of moir{\'e} superlattices},
  author  = {Duncan, Cameron J. R. and Johnson, Amalya C. and Maity, Indrajit and Rubio, Angel and others},
  journal = {Nature},
  volume  = {647},
  pages   = {619--624},
  year    = {2025},
  doi     = {10.1038/s41586-025-09707-3},
}

@article{vanOudheusden2010,
  title={Compression of subrelativistic space-charge-dominated electron bunches for single-shot femtosecond electron diffraction},
  author={Van Oudheusden, T and Pasmans, PLEM and Van Der Geer, SB and De Loos, MJ and Van Der Wiel, MJ and Luiten, OJ},
  journal={Physical Review Letters},
  volume={105},
  number={26},
  pages={264801},
  year={2010},
  publisher={APS}
}

@article{Kealhofer2016,
  title={All-optical control and metrology of electron pulses},
  author={Kealhofer, Catherine and Schneider, Waldemar and Ehberger, Dominik and Ryabov, Andrey and Krausz, Ferenc and Baum, Peter},
  journal={Science},
  volume={352},
  number={6284},
  pages={429--433},
  year={2016},
  publisher={American Association for the Advancement of Science}
}

@article{Qi2020,
  title={Breaking 50 femtosecond resolution barrier in MeV ultrafast electron diffraction with a double bend achromat compressor},
  author={Qi, Fengfeng and Ma, Zhuoran and Zhao, Lingrong and Cheng, Yun and Jiang, Wenxiang and Lu, Chao and Jiang, Tao and Qian, Dong and Wang, Zhe and Zhang, Wentao and others},
  journal={Physical Review Letters},
  volume={124},
  number={13},
  pages={134803},
  year={2020},
  publisher={APS}
}

@article{Yang2025,
  title={Sub-5-fs compression and synchronization of relativistic electron bunches enabled by a high-gradient $$\backslash$alpha $-magnet and low-jitter photoinjector},
  author={Yang, Yining and Wang, Zhiyuan and Lv, Peng and Song, Baiting and Huang, Pengwei and Jia, Yanqing and Liu, Zhuoxuan and Zheng, Lianmin and Huang, Wenhui and Musumeci, Pietro and others},
  journal={arXiv preprint arXiv:2508.03946},
  year={2025}
}

@Article{Ryabov2025,
  title  = {High-repetition-rate terahertz and ultraviolet radiation for high-throughput ultrafast electron diffraction},
  author = {Ryabov, A. and Amini, K.},
  journal= {arXiv},
  volume = {X},
  number = {X},
  pages  = {X},
  year   = {2025},
  publisher = {arXiv},
  platform    = {ArXiv:2507.23727},
  doi    = {XYZ},
  url    = {https://doi.org/XYZ},
}

@article{Otto2017,
  title={Solving the jitter problem in microwave compressed ultrafast electron diffraction instruments: Robust sub-50 fs cavity-laser phase stabilization},
  author={Otto, Martin R and Ren{\'e} de Cotret, LP and Stern, Mark J and Siwick, Bradley J},
  journal={Structural Dynamics},
  volume={4},
  number={5},
  year={2017},
  publisher={AIP Publishing}
}

@article{Centurion2022,
  title={Ultrafast imaging of molecules with electron diffraction},
  author={Centurion, Martin and Wolf, Thomas JA and Yang, Jie},
  journal={Annual Review of Physical Chemistry},
  volume={73},
  pages={21--42},
  year={2022},
  publisher={Annual Reviews}
}

@Book{Amini2023,
title     = {Structural Dynamics with X-ray and Electron Scattering},
publisher = {Royal Society of Chemistry},
year      = {2023},
author    = {Amini, Kasra and Rouzée, Arnaud and Vrakking, J. J. Marc},
isbn      = {978-1-83767-114-4},
doi       = {10.1039/9781837671564},
}

@article{Yang2018,
  title={Imaging CF$_3$I conical intersection and photodissociation dynamics with ultrafast electron diffraction},
  author={Yang, Jie and Zhu, Xiaolei and Wolf, Thomas JA and Li, Zheng and Nunes, J Pedro F and Coffee, Ryan and Cryan, James P and G{\"u}hr, Markus and Hegazy, Kareem and Heinz, Tony F and others},
  journal={Science},
  volume={361},
  number={6397},
  pages={64--67},
  year={2018},
  publisher={American Association for the Advancement of Science}
}

@article{Wolf2019,
  title={The photochemical ring-opening of 1, 3-cyclohexadiene imaged by ultrafast electron diffraction},
  author={Wolf, Thomas JA and Sanchez, David M and Yang, J and Parrish, RM and Nunes, JPF and Centurion, M and Coffee, R and Cryan, JP and G{\"u}hr, Markus and Hegazy, Kareem and others},
  journal={Nature Chemistry},
  volume={11},
  number={6},
  pages={504--509},
  year={2019},
  publisher={Nature Publishing Group UK London}
}

@article{Yang2020,
  title={Simultaneous observation of nuclear and electronic dynamics by ultrafast electron diffraction},
  author={Yang, Jie and Zhu, Xiaolei and F. Nunes, J Pedro and Yu, Jimmy K and Parrish, Robert M and Wolf, Thomas JA and Centurion, Martin and G{\"u}hr, Markus and Li, Renkai and Liu, Yusong and others},
  journal={Science},
  volume={368},
  number={6493},
  pages={885--889},
  year={2020},
  publisher={American Association for the Advancement of Science}
}

@article{Jiang2025,
  title={Super-resolution femtosecond electron diffraction reveals electronic and nuclear dynamics at conical intersections},
  author={Jiang, Hui and Zhang, Juanjuan and Wang, Tianyu and Peng, Jiawei and Jin, Cheng and Zou, Xiao and Zhu, Pengfei and Jiang, Tao and Lan, Zhenggang and Yong, Haiwang and others},
  journal={Nature Communications},
  volume={16},
  number={1},
  pages={6703},
  year={2025},
  publisher={Nature Publishing Group UK London}
}

@article{Wang2025,
  title={Imaging the photochemical dynamics of cyclobutanone with MeV ultrafast electron diffraction},
  author={Wang, Tianyu and Jiang, Hui and Jin, Cheng and Zou, Xiao and Zhu, Pengfei and Jiang, Tao and He, Feng and Xiang, Dao},
  journal={Journal of Chemical Physics},
  volume={162},
  number={18},
  year={2025},
  publisher={AIP Publishing}
}

@article{Wang2025a,
  title={Probing valence electron and hydrogen dynamics using charge-pair imaging with ultrafast electron diffraction},
  author={Wang, Tianyu and Jiang, Hui and Zhang, Ming and Zou, Xiao and Zhu, Pengfei and He, Feng and Li, Zheng and Xiang, Dao},
  journal={arXiv preprint arXiv:2506.21047},
  year={2025}
}

@article{FernandezPerez2021,
  title={Characterization of a hybrid pixel counting detector using a silicon sensor and the IBEX readout ASIC for electron detection},
  author={Fernandez-Perez, S and Boccone, V and Broennimann, C and Disch, C and Piazza, L and Radicci, V and Rissi, M and Schulze-Briese, C and Trueb, P and Zambon, P},
  journal={Journal of Instrumentation},
  volume={16},
  number={10},
  pages={P10034},
  year={2021},
  publisher={IOP Publishing}
}

@article{Kealhofer2015,
  title={Signal-to-noise in femtosecond electron diffraction},
  author={Kealhofer, Catherine and Lahme, Stefan and Urban, Theresa and Baum, Peter},
  journal={Ultramicroscopy},
  volume={159},
  pages={19--25},
  year={2015},
  publisher={Elsevier}
}


\end{document}